\documentclass[journal]{IEEEtran}
\usepackage{multicol}
\usepackage{cite}
\usepackage[dvips]{graphicx}
\usepackage{epsfig}
\graphicspath{{Figures/}}
\usepackage{subfigure}
\usepackage{float} 
\usepackage{gensymb}
\usepackage{array}
\usepackage{amsmath}
\usepackage{amsthm}
\usepackage{algorithm}
\usepackage{algorithmic}
\usepackage{mdwmath}
\usepackage{mdwtab}
\usepackage{amsfonts}
\usepackage{psfrag}
\usepackage{textcomp}
\usepackage{psfrag}
\usepackage{color}

\newtheorem*{proof*}{Proof}

\usepackage{url}
\begin{document}
\title{Association, Blockage and Handoffs in IEEE 802.11ad based 60\,GHz Picocells- A Closer Look}
\author {\IEEEauthorblockN{Kishor Chandra Joshi, Rizqi Hersyandika and R. Venkatesha Prasad}\\
\thanks{Kishor Chandra Joshi is with CNRS/CentraleSupelec, University of Paris-Saclay, Paris, France and TU Delft, Netherlands. } 
\thanks{R. Hersyandika is with Spectrum Monitoring department, MCIT, Indonesia} 
\thanks{R. Venkatesha Prasad is with TU Delft, Netherlands}  }
\maketitle
\begin{abstract}
  The link misalignment  and high susceptibility to blockages are the biggest hurdles in realizing 60\,GHz based wireless local area networks (WLANs). However, much of the previous  studies  investigating 60\,GHz alignment and blockage issues do not provide an accurate quantitative evaluation from the perspective of WLANs. In this paper, we present an in-depth quantitative evaluation of commodity IEEE 802.11ad devices by forming a 60\,GHz WLAN with two docking stations mimicking as access points~(APs). Through extensive experiments, we provide important insights about directional coverage pattern of antennas, communication range and co-channel interference and blockages. We are able to measure the IEEE 802.11ad link alignment and association overheads in absolute time units.  With a very high accuracy (96-97\%), our blockage characterization can differentiate between temporary and permanent blockages caused by humans in the indoor environment, which is a key insight. Utilizing our blockage characterization, we also demonstrate intelligent handoff to alternate APs using consumer-grade IEEE 802.11ad devices. Our blockage-induced handoff experiments  provide important insights that would be helpful in integrating millimeter wave  based WLANs into future wireless networks.  
\end{abstract}

\section{Introduction}
Due to the availability of a large bandwidth in the millimeter wave (mmWave) frequency band (30\,GHz to 300\,GHz), it has become a key enabler for providing multi-Gbps wireless access in future 5G-and-beyond networks~\cite{mmwaveMobilenetworks,cogcel}. In particular, the 60\,GHz band has been of special interest for high speed wireless local area networks (WLANs)~\cite{standardization60GHz}. Comparing with the 2.4/5\,GHz bands, mmWave propagation is subjected to a very high free-space path loss. For example, path loss at 60\,GHz is at least 20\,dB more than that at 5\,GHz~\cite{cogcel}.  Directional antennas are used to compensate for the high path loss.  Another important propagation characteristic of mmWave signals is the susceptibility to blockage by obstacles, e.g., the human blockage can result in an attenuation up to 20\,dB~\cite{Jacob2011}. 
 
 The existing 60\,GHz commercial off-the-shelf (COTS) devices have mainly employed WiGig (IEEE 802.11ad) specifications~\cite{80211ad}. The literature on the experimental evaluations of IEEE 802.11ad based devices is limited and mainly focused on the isolated link characterization measuring throughput, packet structures and received signal strength, attenuation due to obstacles, etc. For example, \cite{demystify60GHz} and \cite{saha2} have used Wilicity's IEEE 802.11ad hardware to investigate the throughput, communication range, blockage-induced attenuation and beam-steering capabilities in outdoor and indoor environments, respectively. Similarly, \cite{Nitsche} also uses Wilocity's IEEE 802.11ad hardware and provides important insights on the imperfections of antenna radiation patterns, interference caused by side lobes and the dynamics of frame aggregation mechanism. There are very few works~\cite{FlexbeamWisconsin, Sur2016, zhang2016openmili, hassanieh2017agile}  providing important insights on the overheads related to beam-searching, misalignments, and beamwidth adaptations. However, use of customized hardware and software makes it difficult to benchmark the characteristics of commodity IEEE 802.11ad systems. Thus, there is little understanding of how the special characteristics of 60\,GHz systems (e.g., directionality, blockage by obstacles, beam alignment) affect the prospect of providing seamless WLAN and cellular communication. 
 
 The \textit{major gaps} that exist vis-\`a-vis the evaluation of commodity IEEE 802.11ad devices are as follows:  (i)~Throughput degradation due to the antenna alignment is investigated but an actual assessment of rebeamforming in absolute time units is missing. (ii)~The spatial reuse capacity of 60\,GHz links is highlighted in the literature but the impact of 'deafness' arising due to the inability of narrow beam antennas to sense each others' transmissions in IEEE 802.11ad has not received much attention, except in ~\cite{Nitsche}. (iii)~Link blockage investigations are limited to the measurement of attenuation in received signal quality, but the link behavior for different types of blockages (i.e., temporary and permanent) and their impact on providing seamless WLAN/cellular experience is not pursued. (iv)~Experimental investigations on the handoff in 60\,GHz networks with multiple access points(AP)/base stations (BS) employing COTS IEEE 802.11ad devices has not been attempted so far. 

Although the first generation of COTS 60\,GHz devices were mainly developed as a replacement of HDMI cables, a wide adoption of 60\,GHz technology for WLANs and mobile communications is imminent in the future 5G-and-beyond wireless systems. Therefore, in this paper, we evaluate the performance of IEEE 802.11ad devices from a WLANs/picocell perspective aiming to fill the above listed gaps. We use INTEL's IEEE 802.11ad chipsets instead of the Wilicity chipsets used in \cite{demystify60GHz,saha2,Nitsche}. Our carefully designed experimental scenarios coupled with extensive measurements provide important insights on the misalignment overhead, device discovery/association times, the impact of human blockage and presence of alternate non-line-of-sight (NLOS) path on the 60\,GHz links. We summarize here our main contributions:\newline
(1)~Using extensive measurements, we provide for the first time, an actual estimate of the device association and alignment times in IEEE 802.11ad systems.\newline
(2)~We show that the deafness arising due to the use of directional antennas in 60\,GHz communications can heavily deteriorate the link performance.\newline
(3)~We propose a novel human blockage characterization mechanism that differentiates -- with a very high accuracy (up to 96-100\%) -- the different types of human blockages.\newline
(4)~This is the first work that creates and studies an IEEE 802.11ad based WLAN system with multiple APs and evaluates the blockage and mobility triggered handoff performance. 

The rest of the paper is organized as follows. In Section~\ref{sec:RW} we discuss related works. Section~\ref{sec:Prelims} provides the basic characteristics of used COTS devices. We investigate the beam alignment and association overheads in Section~\ref{sec:association} which is followed by co-channel intereference characterization in Section~\ref{sec:deafness}. The human blockage characterization, blockage induced handoffs and  mobility induced handoffs are presented in Section~\ref{sec:humanblockage}. Section~\ref{sec:conclusion} concludes the paper.
 \section{Related Works}\label{sec:RW}
Performance evaluation of commercially available IEEE 802.11ad based systems has received much attention in recent years. Feasibility of 60\,GHz outdoor picocells employing Wilocity's 802.11ad chipsets is explored in \cite{demystify60GHz}. Authors show that it is indeed feasible to use 60\,GHz band for high data rate communications in outdoor picocells despite the high path loss, susceptibility to blockages by obstacles, and movements induced beam switching resulting in link outages. Investigation on throughput and communication range of 60\,GHz radios using Wilocity IEEE 802.11ad chipset is presented in~\cite{Saha1} and \cite{saha2}. It is argued in \cite{Saha1} that relays can significantly improve the performance of IEEE 802.11ad systems. 

Ansari et al.,~\cite{Petri_Mahonen} have measured the bit error rates and throughput performance of various 60\,GHz transceivers which are used for empirical characterization of 60\,GHz links. Nitsche et al.,~\cite{Nitsche} provide important insights into the frame level protocol analysis of 60\,GHz links using Dell D5000 (which employs Wilocity chips) and WiHD-compatible DVDO Air-3c system as transmitters and Vubiq 60\,GHz system receiver to capture the over-the-air transmissions. It is shown that side lobes can result in significant interference up to 5\,m distance. It is shown that even a small link misalignment can significantly degrade the throughput neighboring links. Using Dell D5000 docking stations and Dell Lattitude E7440 (using Wilocity WiGig chipsets), Loch et al.,~\cite{Loch2016a} propose to use frame aggregation to counter the frame loss due to collisions when multiple 60\,GHz links operate simultaneously. In \cite{palacios2016tracking}, fast beam training and tracking mechanisms are proposed by employing hybrid analog-digital transceivers to simultaneously collect the channel information from multiple directions. 

Antenna misalignment and beam searching overheads have extremely adverse effects on the performance of 60\,GHz links and can lead to a significant degradation in the link throughput~\cite{FlexbeamWisconsin}. Simic et al.~\cite{Simic2016} performed measurements in indoor and outdoor environments and argued that misalignment can result in frequent outages.  Using customized programmable 60\,GHz radio platform called WiMi, Sur et al.~\cite{Sur2016} propose BeamSpy, an algorithm to predict the quality of alternative beams to reduces beam searching overhead. Zhang et al.~\cite{zhang2016openmili} have proposed LABA: a learning assisted beam-adaptation mechanism to minimize the beam searching overhead, which is implemented on OpenMili, the next generation of WiMi.   Haitham et al.~\cite{hassanieh2017agile} have implemented a beam alignment protocol on customized 60\,GHz software-defined radio platform that uses sparse-recovery theory to minimize the beam-search space and results in fast beam switching. Haidar et al.~\cite{edward} have proposed MOCA, that invokes beam sounding  before each time a packet is transmitted to estimate the link quality of selected beams, to increase the link mobility resiliency. In \cite{fsaha2018fast}, link blockage evaulation considering mobile and static blockages is presented, however, its impact on handoffs is ignored. 

The works listed above have three major limitations: Firstly, most of these papers investigate the performance of 60\,GHz systems from the perspective of a single link. On the other hand, we investigate the IEEE 802.11ad systems from a dense WLANs or picocellular system perspective where frequent handoffs can be a major challenge due to the mobility or blockages. Secondly, previous works on human blockage are merely limited to finding how much signal attenuation is caused by a blocking person. However, we analyze the dynamics of human shadowing in detail which is highly important in facilitating the handoffs in 60\,GHz dense WLANs and picocellular networks. Thirdly, an accurate estimate of alignment time is missing in the literature for the COTS IEEE 802.11ad devices, while we provide accurate estimates of beam alignment and initial user association times for IEEE 802.11ad COTS chipsets.
 \section{Preliminary System Characterization}\label{sec:Prelims}
   We use INTEL WiGig sink chipset W13100~\cite{Intel_w13100} as the AP and a Dell laptop equipped with INTEL tri-band wireless-AC 17265~\cite{Intel_ac17265} wireless card as the wireless station (STA). It uses IEEE 802.11ad Control PHY (MCS\,0) for control frame transmissions and Single Carrier PHY (MCS\,1-12) for the data transmission supporting the data rate in the range of 385--4620\,Mbps. An application programming interface (API) provides the normalized signal quality parameter in the scale of 0 to 10 which indicates the link quality. The lower layer (PHY and MAC) information is not exposed to users, which is the biggest limitation of all commercially available first generation IEEE 802.11ad chipsets~\cite{saha2,Nitsche,demystify60GHz}.   
  \begin{figure}[!]
  	\centering 
x  	\includegraphics[width = 0.20\textwidth]{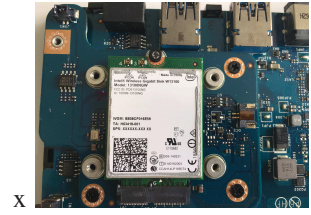}
  	\caption{INTEL's WiGig sink chipset W13100.}
  	\label{FigWiGigchips}	
  	\vspace{-4mm}
  \end{figure}
  In our experiments we primarily depend on two parameters: (i)~normalized signal quality provided by the WiGig API; and (ii)~throughput measured using the iPerf client~\cite{iperf}. 
  
  We represent the signal quality parameter with $q$, where $q\in{[0,10]}$. A Dell Latitude-E6430 hosting iPerf server generates the TCP traffic and is connected to the  WiGig sink chipset W13100 via Gigabit Ethernet interface. This connection limits the maximum data rate to 1\,Gbps due to the limitation of Gigabit Ethernet port used.   Despite the limitations of the experimental setup, we provide powerful insights on the behavior of 60\,GHz WLAN/picocellular network with extensive experiments that are carefully designed to catch the key characteristics of mmWave links in a WLAN scenario.  
  \subsection{Directional coverage pattern}
  The first step in characterizing the WiGig chipsets is measuring the coverage patterns of the 2$\times$ 8 antenna array embed in the INTEL's chipsets. Initially, both the AP and STA are positioned 4\,m apart in such a way that the antenna array modules face each other ensuring line-of-sight (LOS) connection.  To measure the directional coverage in azimuth plane, one device is rotated 360\textdegree\, in a step of 10\textdegree\, while the other is kept fixed and the signal quality 'q' is recorded at each 10\textdegree\, steps. 
  
\begin{figure}[!]
	\centering 
	\subfigure[Access point's coverage pattern.]{
	\includegraphics[width=0.220\textwidth]{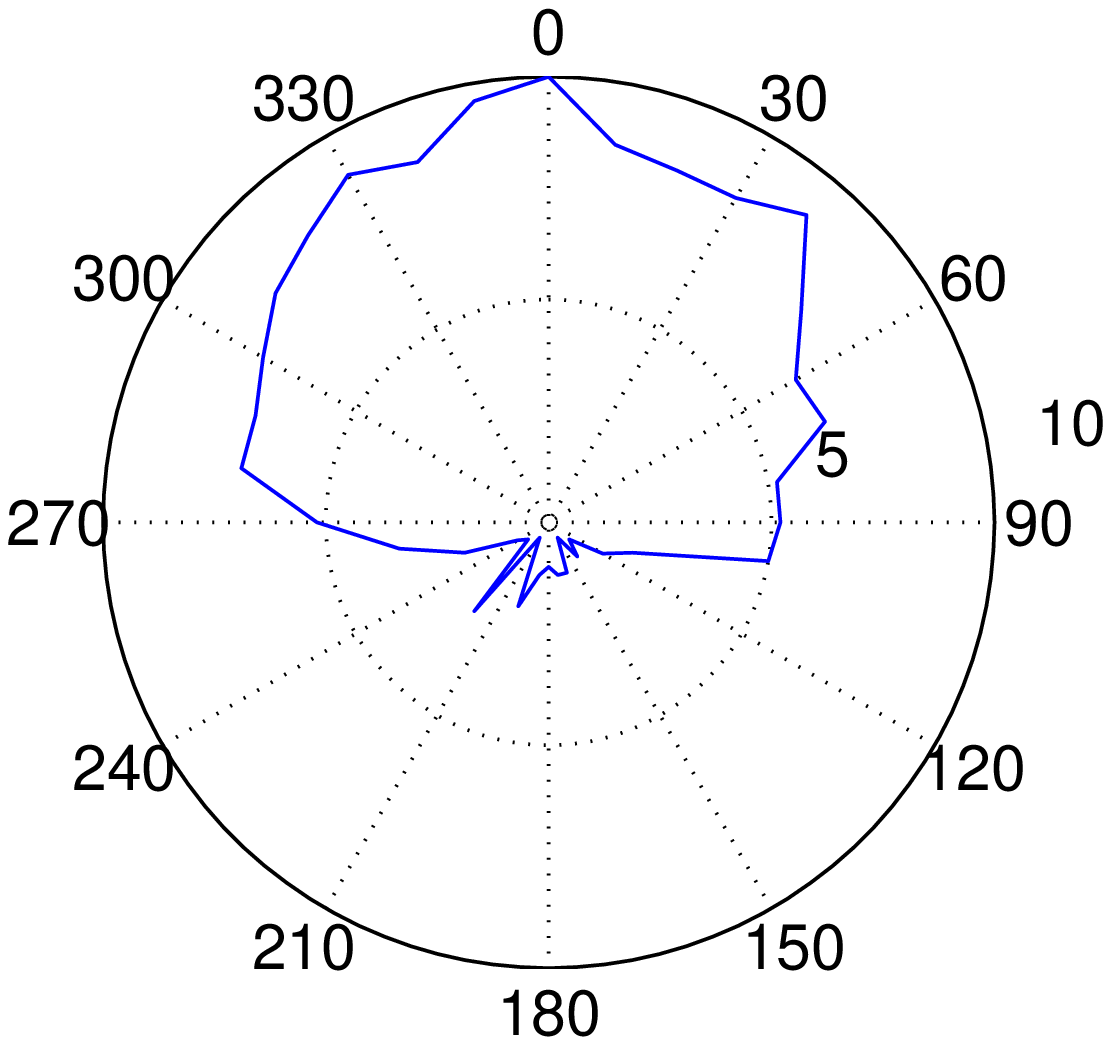}
	\label{fig3:a}
	 }	
	\subfigure[Laptop's  coverage pattern.] {
	\includegraphics[width=0.22\textwidth]{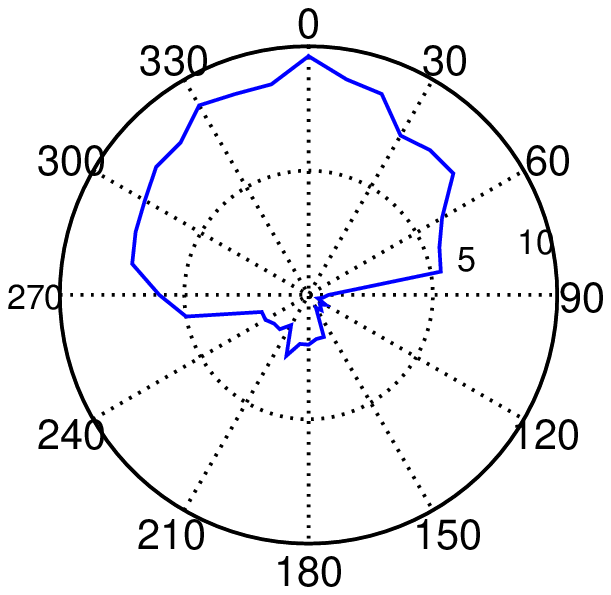}
	\label{fig3:b}
	}
		\caption{Directional (azimuth) coverage measurements.}
	\label{fig3.6:directional_coverage}	
	\vspace{-4mm}
\end{figure}
Fig.~\ref{fig3:a} and Fig.~\ref{fig3:b} show the directional coverage measurements of the AP and the STA, respectively. The results show that antenna arrays of both AP and STA have a non-uniform radiation pattern (antenna gains) in 360\textdegree\, azimuth plane.  It is indicated by the high signal quality monitored within 90{\textdegree\,} to 270{\textdegree\,} angle. This shows the limitations of cost-effective designs used in consumer-grade IEEE 802.11ad chipsets resulting in a skewed coverage in 360\textdegree\, azimuth plane.
\subsection{Communication range}
To measure the communication range performnace (distance) of IEEE 802.11ad AP, we performed experiments in both indoor and outdoor environment. Indoor environment consists of a narrow corridor of width 2\,m with hard concrete walls on both the  sides, while the outdoorenvironment consists of an empty park. There were no obstacle between STA and AP. We considered two cases, namely, (i)~with the data traffic (using iPerf connection); and (ii)~without the data traffic (only control traffic related to beaconing, etc., was present). As the distance  from AP increases, the signal quality $q$ received by STA decreases. At a certain distance, STA is disconnected from AP due to the lack of sufficient received signal strength, or data transmission is aborted as the minimum data rate cannot be supported. This distance is referred  as the maximum communication range.\normalcolor
\begin{figure}[b]
	\centering
\subfigure[Communication range.]{
	\includegraphics[width=0.22\textwidth]{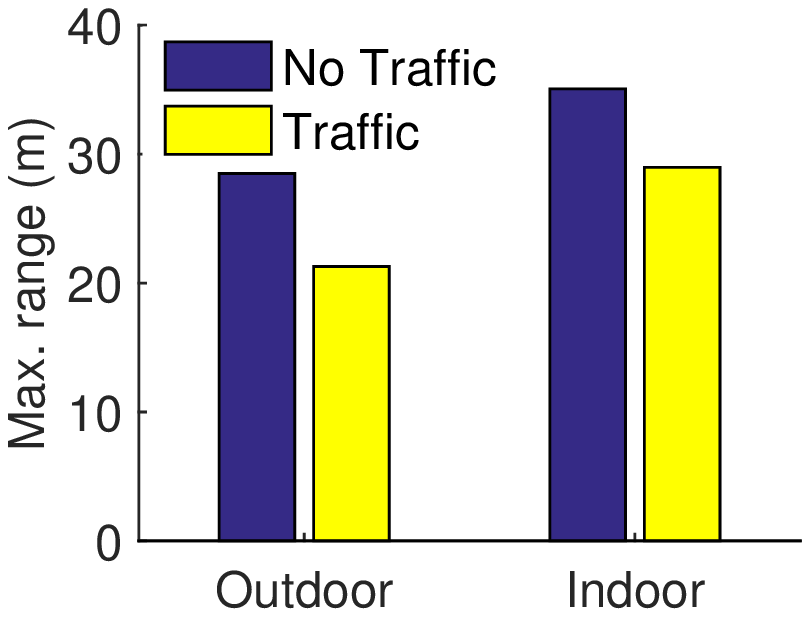}
		\label{fig3.7:maxrange}
	}
	\subfigure[Data rates vs distance in indoors.]{
		\includegraphics[width=0.22\textwidth]{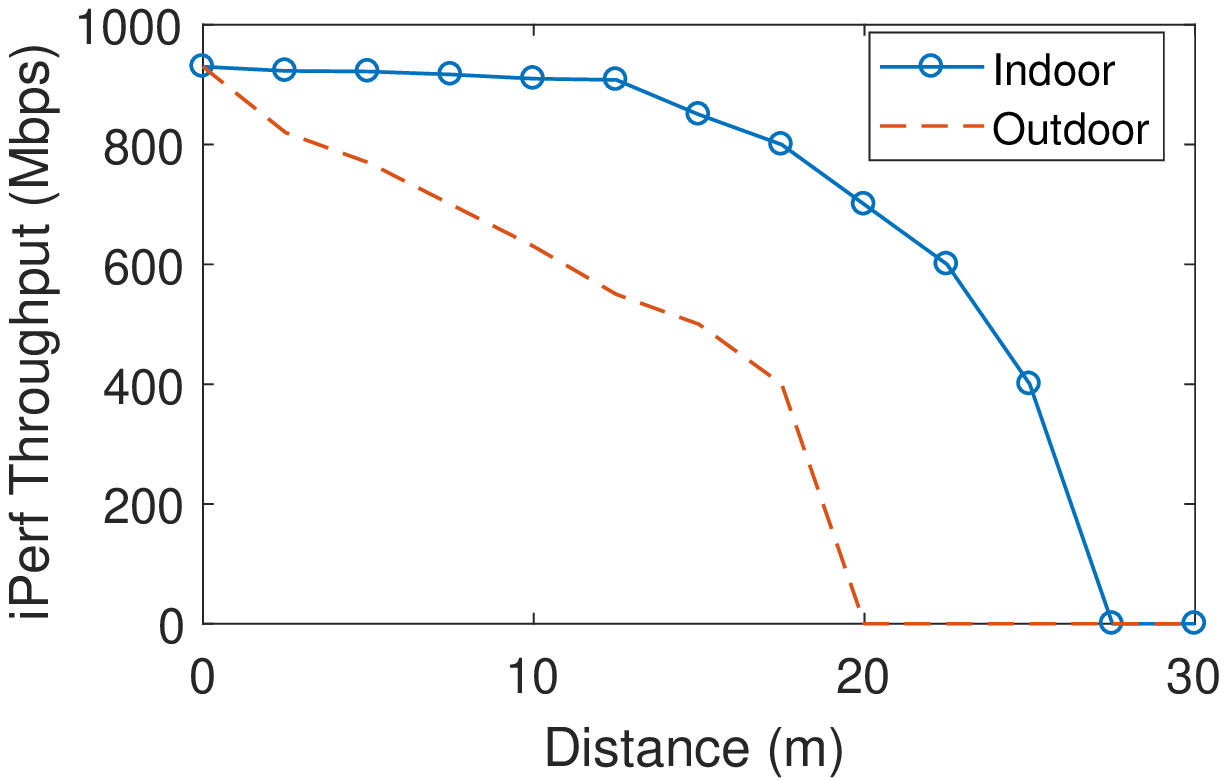}
			\label{fig3.7datarate_distance}
		}
	\caption{Maximum communication range and data rates.}
\end{figure}
 Fig.~\ref{fig3.7:maxrange} shows that the  maximum range obtained in indoor environment is higher than that in the outdoor environment. Since indoor environment consists of narrow corridor,  the reflections through the  corridor walls reinforce the LOS component which boosts the communication range.  Fig.~\ref{fig3.7datarate_distance} also shows the data rates in indoor (corridor). In each environment, the maximum range is always greater in the absence of data traffic because only the control PHY (MCS0) is used in this case. On the other hand, when data traffic is present, higher MCSs are used that requires greater received signal levels to maintain the connectivity.\normalcolor
\section{Association and Beam Alignment Overheads}\label{sec:association}
\subsection{Association}
 Initially, when the STA is not yet connected to the AP, the measured signal quality $q$ is zero. After the STA receives the first beacon, the association process begins.  We measured the time difference between  the first registered signal value and the final stable signal value, which is defined as association time.\normalcolor
\begin{figure}[!]
	\centering
	\subfigure[Association time.]{
	\includegraphics[width=0.20\textwidth]{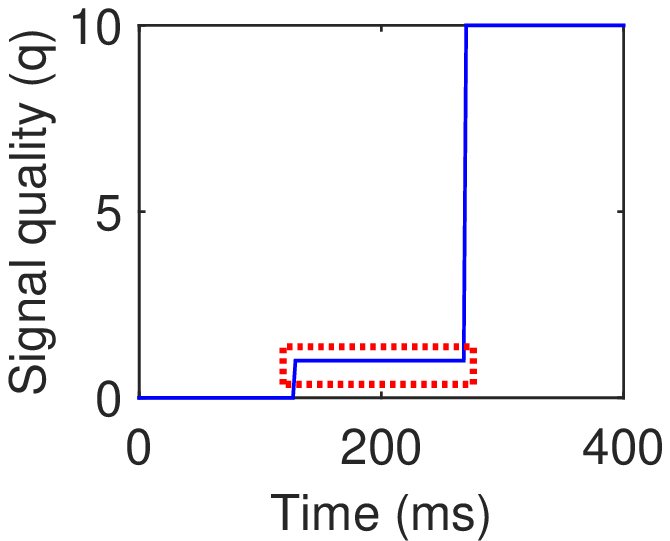}
		\label{fig3.4:association_time}}
\subfigure[Beam re-alignment.]{
	\includegraphics[width=0.20\textwidth]{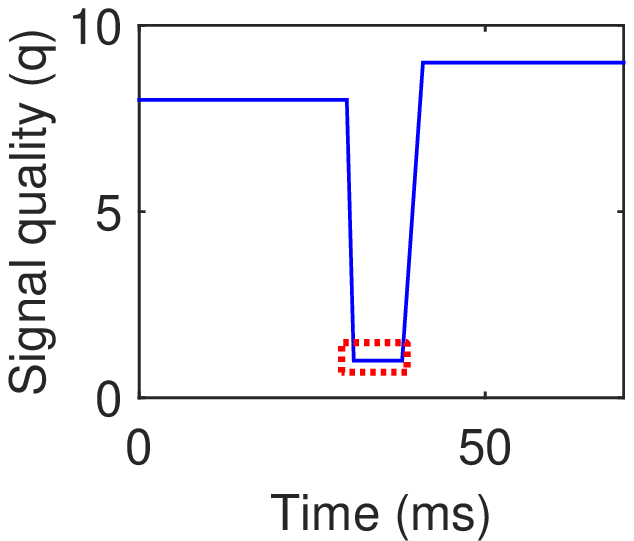}
		\label{fig3.5:beam_realignment_time}}
	\caption{A snapshot of signal quality variation during association and beam re-alignment.}
	\label{association_and_alignment}
	\vspace{-4mm}
\end{figure}

An example of signal quality $q$ during the association process is depicted in Fig.~\ref{fig3.4:association_time}.
\begin{table}[!]
\small
\caption{IEEE 802.11ad user association time.}
\label{table_associationTime}
\centering
  \begin{tabular}{|c|c||c|c|}
  \hline
  Tx/Rx Distance (m) & 1 & 4\ & 10 \\
  \hline
  Best case (ms) & 146.28 & 153.34& 172.12\\
  \hline
  Worst case (ms) & 386.14 & 375.56 & 324.58 \\
  \hline
  Average (ms) & 247.47 & 245.4 & 243.75\\
  \hline
  \end{tabular}
  \vspace{-4mm}
   \end{table}
In Table~\ref{table_associationTime} we show the association time for different distances -- 1\,m, 4\,m and 10\,m. The measurement results show that the average association time is around 240\,ms. 20 repetitions were performed for each AP-STA distance value. The average association time for all the three distances is almost equal which is logical because IEEE 802.11ad uses two-stage fixed beamwidth searching mechanism. 
\subsection{Beam re-alignment time}
The use of directional antennas makes IEEE 802.11ad links  highly susceptible to misalignment resulting from change in the orientation of devices or use movements. It is natural that if misalignment happens, device-pair tries to realign antenna beams to restore the link quality. We define, the beam realignment time as the duration required by an STA and an AP to realign their beam when the alignment is disturbed. In our experiments, AP and STA were kept 1\,m above the ground with a LOS connection. We rotate AP  in steps of 60\textdegree.  Fig.~\ref{fig3.5:beam_realignment_time} illustrate an example of signal quality snapshot during the realignment duration.\normalcolor We can see that immediately after rotation, STA and AP try to realign their beams in order to retrieve highest achievable signal quality $q$. The average beam realignment time (of 20 repetition) is 7.65\,ms. As opposed to existing notion that re-alignment can take too much time, this is a significant result  showing the capability of COTS devices to quickly find the alternate beams.
\section{Deafness and Interference}\label{sec:deafness}
Generally, directional antennas should cause less interference to neighboring links due to the confined transmissions. However, they are highly prone to the problem of ``deafness’’ in which transmitter of one link becomes unaware of the existing other transmission resulting in frame collisions. This is particularly a major challenge when carrier sensing based MAC protocols are used. 
\begin{figure}[!]
	\centering
	\subfigure[Interfering links with deafness.]{
	\includegraphics[width=0.35\textwidth]{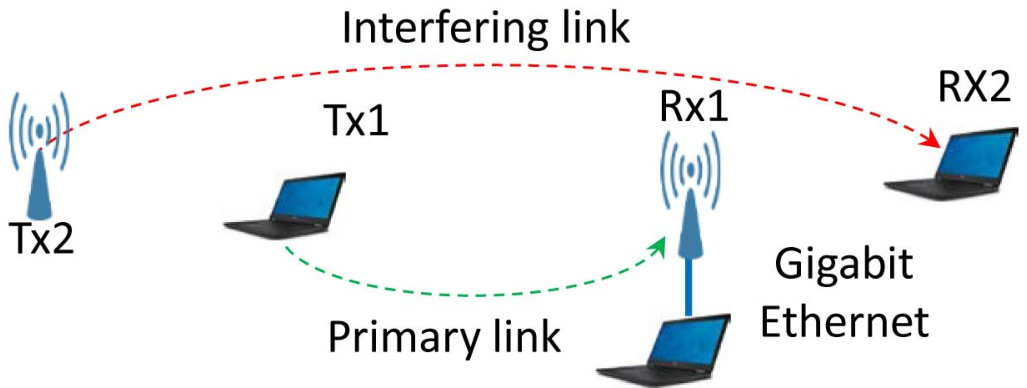}
		\label{fig3.8:interference_sc1}}
\subfigure[Interfering links able to sense mutual transmissions.]{
	\includegraphics[width=0.35\textwidth]{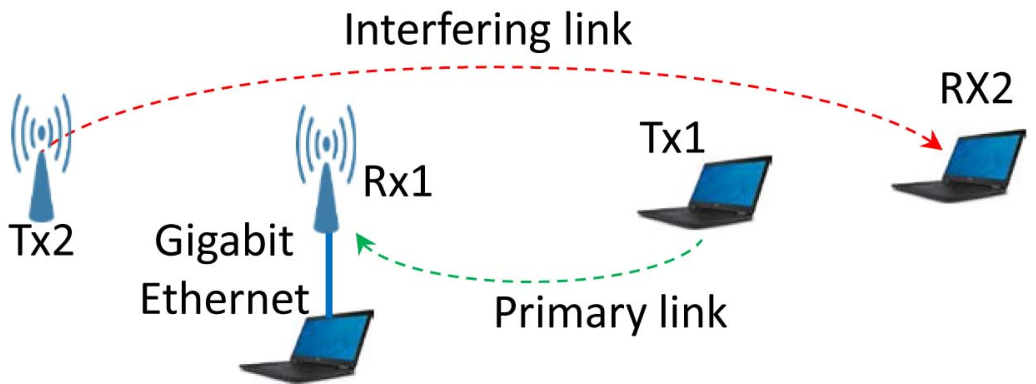}
	\label{fig3.9:interference_sc2}}
		\label{cochannel}
			\caption{Scenarios illustrating multiple IEEE 802.11ad links.}
			\vspace{-3mm}
\end{figure}
To assess the impact of deafness, we created a measurement setup having two scenarios where two IEEE 802.11ad  links (each link consists of an AP and a STA) simultaneously operate as depicted in Fig.~\ref{fig3.8:interference_sc1} and Fig.~\ref{fig3.9:interference_sc2}. The distance between successive devices in both the scenarios is 1\,m .
  
In the first scenario (Fig.~\ref{fig3.8:interference_sc1})  interfering transmitter is not able to sense the presence of primary transmitter thus creating a deaf node. In the second scenario (Fig.~\ref{fig3.9:interference_sc2}.), transmitters of the primary and interfering link are facing each other, and hence both are aware of each other.  Both the links use the same channel. The traffic in the primary connection was generated using iPerf3. The traffic in the interfering connection was produced by the a large file transfer from the  AP to STA. 

\begin{figure}[!]
	\centering 
	\subfigure[Scenario-1.] 
	{\includegraphics[width=0.21\textwidth]{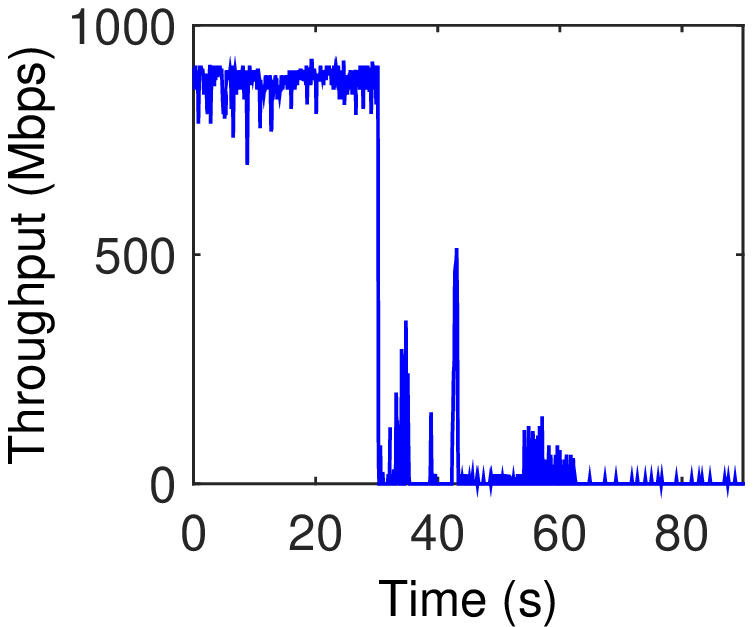}}	
	\subfigure[Scenario-2.] 
	{\includegraphics[width=0.21\textwidth]{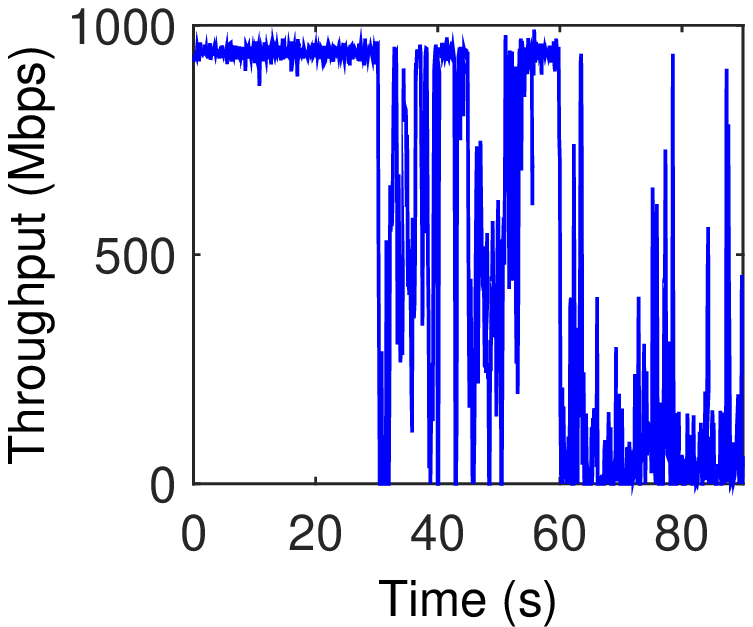}}
	
	\caption{Throughput performance due to co-channel interference.}
	\label{fig3.10:interference_throughput}	
	\vspace{-4mm}
\end{figure}

Fig.~\ref{fig3.10:interference_throughput} show the effect of interfering link on the throughput performance of the primary link in Scenario-1 and Scenario-2, respectively. During the first 30\,s, the interfering link was kept inactive active as indicated by the stable throughput of the primary link. At $t$ = 30\,s, STA1 started copying large files from  AP2. In Scenario-1, the collisions due to deafness result in a significant  throughput degradation. The throughput  remains significantly low for most of the time and cannot recover to the level before the interference was introduced. This deafness effect is a result of the directional transmission in 60\,GHz. Meanwhile, in Scenario-2, the throughput degradation is not so significant which implies that both the primary and  and secondary transmitters are able to sense each other's transmissions and therefore the primary link is able to coexist with the interfering link.\normalcolor
%
\section{Effects of Human Blockage}\label{sec:humanblockage}
Due to small wavelength and narrow beams, 60\,GHz links suffer heavily due to blockages by obstacles. The human induced shadowing is imminent in indoor environments, hence it is important to understand how the  IEEE 802.11ad  links behave during blockages. We designed three experimental scenarios which are likely to happen in human induced shadowing: (i)~transient human blockage, i.e., a short term blockage caused by a person walking across an IEEE 802.11ad link; (ii)~permanent human blockage without NLOS path, i.e., a long term blockage caused by a human standing between an IEEE 802.11ad link, with no possibility of any reflective path; and (iii)~permanent human blockage with NLOS path availability, i.e., a long term blockage caused by a human standing between an IEEE 802.11ad link and there is a possibility of a reflected path between the STA and AP. Each scenario is repeated 100 times to find statistically stable measurements.
\subsection{Transient human blockage}
The transient blockage is introduced by a person walking across the 60\,GHz link at normal indoor walking speed.  We performed transient blockage experiment for two different distances of $d$=3\,m and $d$=7\,m between the STA and AP. The person walked across the link in between the AP and STA.
\begin{figure}[h]
	\centering 
	\subfigure[Signal quality, d=3\,m.] 
	{\includegraphics[width=0.20\textwidth]{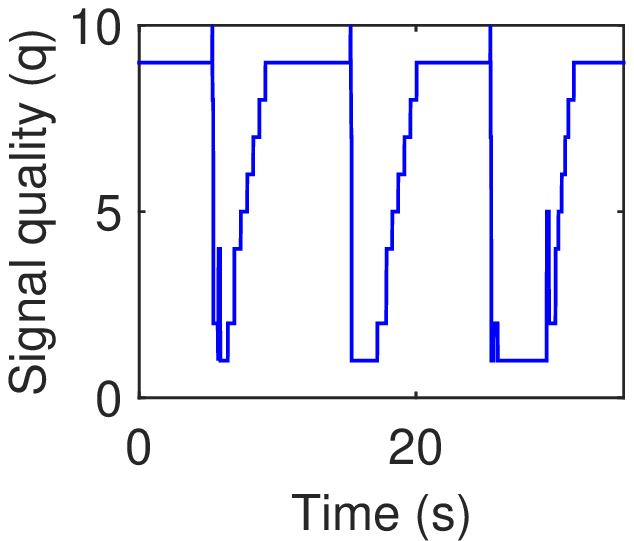}
	\label{fig3.11:scenario1_3m_a}}	
	\subfigure[Throughput, d=3\,m.] 
	{\includegraphics[width=0.20\textwidth]{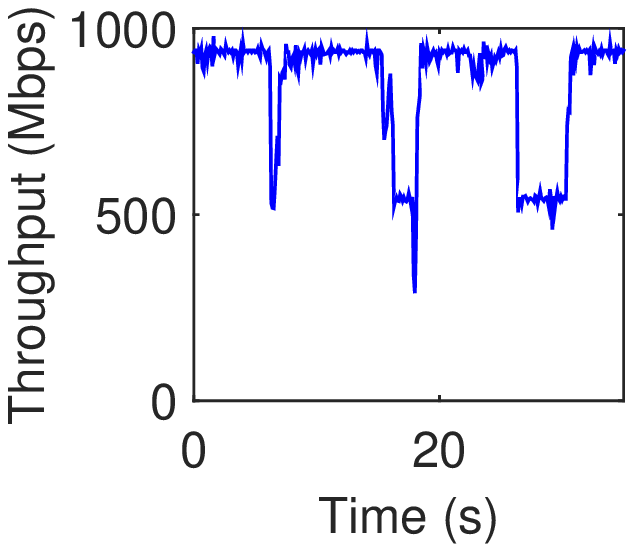}
	\label{fig3.11:scenario1_3m_b}}
	\subfigure[Signal quality, d=7\,m.] 
		{\includegraphics[width=0.20\textwidth]{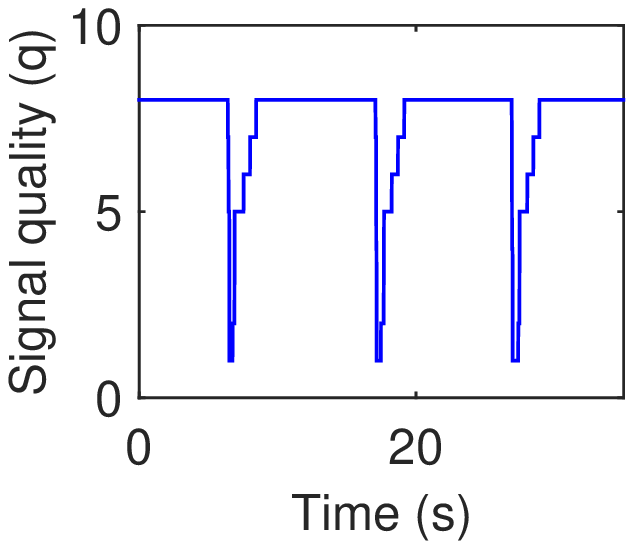}
		\label{fig3.12:scenario1_7m_a}}	
		\subfigure[Throughput, d=7\,m.] 
		{\includegraphics[width=0.20\textwidth]{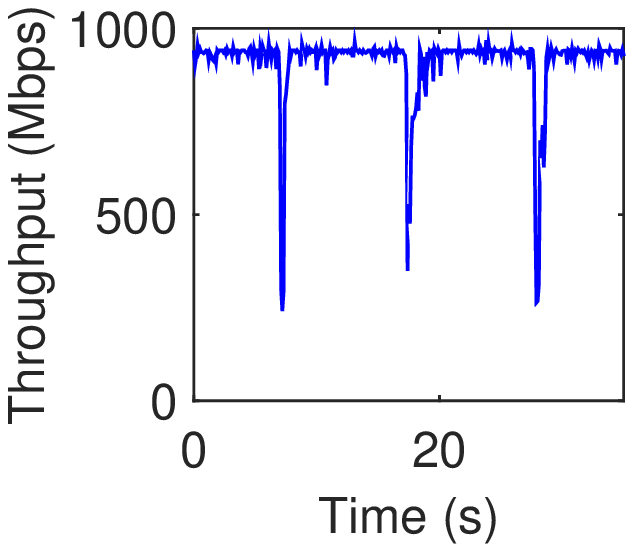}
		\label{fig3.12:scenario1_7m_b}}
	\caption{Transient blockage measurements at $d$ = 3\,m and $d$ = 7\,m.}
	\label{fig3.11:scenario1_3m}
	\vspace{-2mm}	
\end{figure}

%

Fig.~\ref{fig3.11:scenario1_3m_a} and Fig.~\ref{fig3.11:scenario1_3m_b} show the sample results of signal quality $q$ (obtained from WiGig API) and throughput (obtained using iPerf)  when $d$\,=\,3\,m, respectively. 
Three potholes at an interval of approximately 10\,s in both figures indicate the moment when a person crosses the link for three times during 35\,s observation period. During the moments of obstruction, both signal quality  and throughput suffer degradation as indicated by the temporary disruption of both parameters. During these moments, throughput decreases from  900\,Mbps to approximately 500\,Mbps and returns to  900\,Mbps after the obstruction is cleared from the LOS path.

Fig.~\ref{fig3.12:scenario1_7m_a} and  Fig.~\ref{fig3.12:scenario1_7m_b} show the sample results of signal quality $q$ and throughput performance when $d$\,=\,7\,m. Here, average throughput falls from above 900\,Mbps to approximately 250\,Mbps. However, the duration of disruption in the case of $d$ = 7\,m is shorter than that in the case of $d$ = 3\,m although the walking speed was almost same. The reason for a sharper but lesser duration fall in the throughput when $d$ = 7\,m  can be explained as follows. Let us assume the beam is shaped as a cone whose apex is the transmitter. As the distance from the apex (i.e., transmitter) increases, the area of the coverage circle formed by the base of the cone increases proportionally to the square of the distance from the apex. On the other hand, the power density of transmitted signal decreases proportionally to the square of the distance. Thus when $d$ = 3\,m, the coverage circle is smaller, which means the blockage duration is bound to be longer as the probability of a direct path between STA and AP is less compared to the case when $d$ = 7\,m. However, since the power density of transmitted signal is higher when $d$ = 3\,m, the fall in the throughput is less compared to that of  $d$ = 7\,m.
\begin{figure}[!h]
	\centering 
	\includegraphics[width=0.25\textwidth]{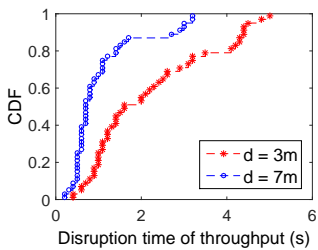}
	\caption{CDF of throughput disruption duration for transient blockage.}
	\label{fig3.13:cdf_time_disruption}	
	\vspace{-3mm}
\end{figure}
We repeated the experiments 50 times for each of the scenarios. Fig.~\ref{fig3.13:cdf_time_disruption} depicts the Cumulative Distribution Function (CDF) of throughput disruption time due to the transient human blockage. The average throughput disruption time in case of $d$ = 3\,m and $d$ = 7\,m are 2.166\,s and 1.036\,s, respectively. The graph confirms that the closer the distance between  STA and AP, the longer is the average shadowing duration due to the human body. 
\subsection{Permanent human blockage without NLOS path}
The permanent blockage is represented by the presence of a human standing in between the STA and AP for a longer duration. Since there is no NLOS path the link suffers a long-term disruption. 

\begin{figure}[!]
	\centering 
	\subfigure[Signal quality, 3\,m.] 
	{\includegraphics[width=0.20\textwidth]{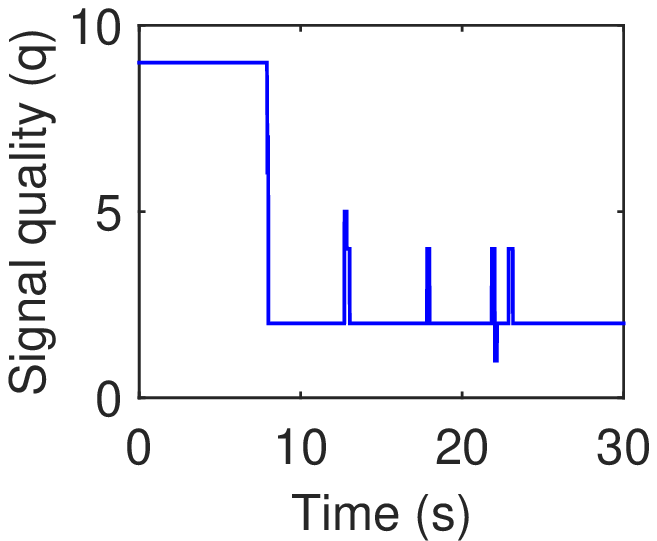}
	\label{fig3.15:scenario2_3m_a}}	
	\subfigure[Throughput, 3\,m.] 
	{\includegraphics[width=0.20\textwidth]{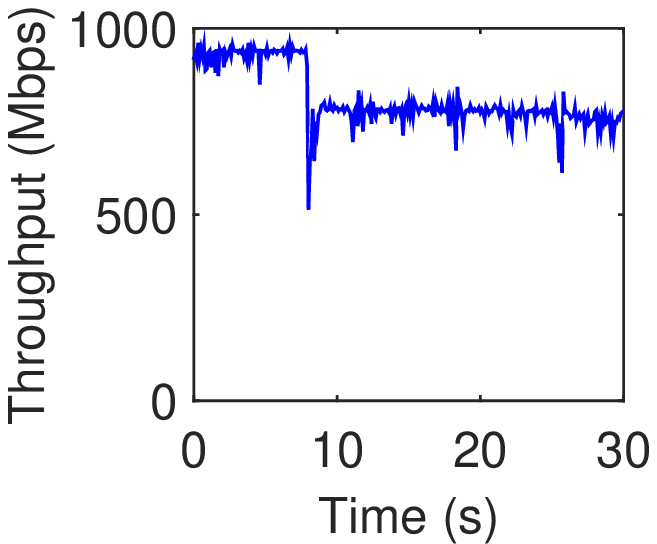}
	\label{fig3.15:scenario2_3m_b}}
	\subfigure[Signal quality, 7\,m.] 
		{\includegraphics[width=0.20\textwidth]{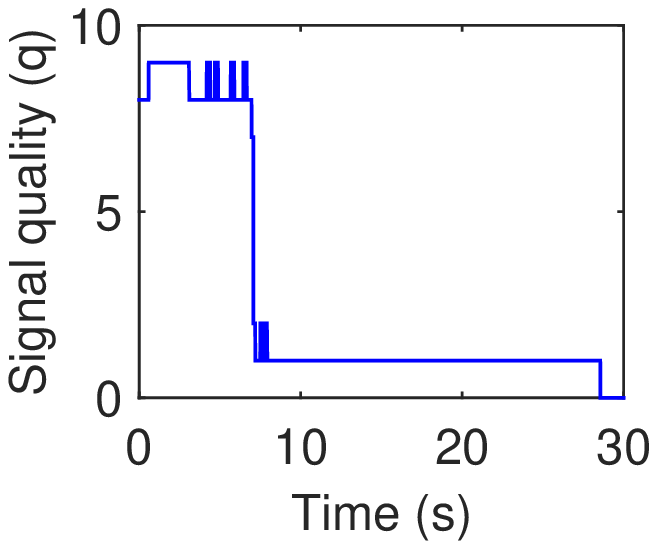}
		\label{fig3.16:scenario2_7m_a}}	
		\subfigure[Throughput, 7\,m.] 
		{\includegraphics[width=0.20\textwidth]{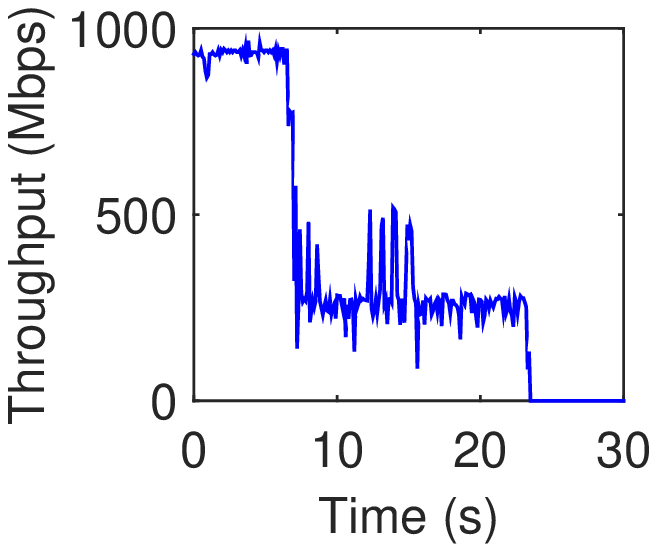}
		\label{fig3.16:scenario2_7m_b}}
	\caption{Permanent blockage measurements, $d$ = 3\,m, $d$ = 7\,m.}
	\label{fig3.15:scenario2_3m}	
	\vspace{-4mm}
\end{figure}

Fig.~\ref{fig3.15:scenario2_3m_a} and  Fig.~\ref{fig3.15:scenario2_3m_b} show the sample result for signal quality $q$ and throughput performance, respectively,  due to the permanent blockage when $d$ = 3\,m. At $t$ = 7\,s, a person starts blocking the link. During the obstruction, the signal quality $q$ falls drastically from $q=9$ to $q=2$. On the other hand, after a sharp dip, the throughput performance is recovered and it manages to achieve the average throughput of approximately 770\,Mbps even during the blockage. A possible reason behind the throughput recovery could be the adaptive MCSs used by IEEE 802.11ad where even a lower order MCS can  provide considerable data rate.
Fig.~\ref{fig3.16:scenario2_7m_a} and Fig.~\ref{fig3.16:scenario2_7m_b} shows the signal quality $q$ and throughput performance, respectively, for $d$ = 7\,m. Here the throughput degradation is more significant compared to the case when $d$ = 3\,m. Initially throughput falls from 900\,Mbps to  250\,Mbps and finally reaches 0\,Mbps at $t$ = 23.4\,s, indicating that the iPerf link is disconnected. The duration when the signal quality $q$ starts decreasing due to the presence of the obstacle until the moment when the link is disconnected is denoted as the disconnection time ($t_{DC}$). From the measurement at $d$ = 7\,m, the average $t_{DC}$ is 16.329\,s. 
 
In general, the permanent blockage causes a persistent degradation in both signal quality $q$ and throughput. As the distance  between AP and the STA increases, this type of blockage can potentially break the link due to the insufficient received signal power leading to high packet loss. 
\subsection{Permanent human blockage with an alternative  NLOS path}
\begin{figure}[!]
	\centering 
	\subfigure[Signal quality, 3\,m] 
	{\includegraphics[width=0.21\textwidth]{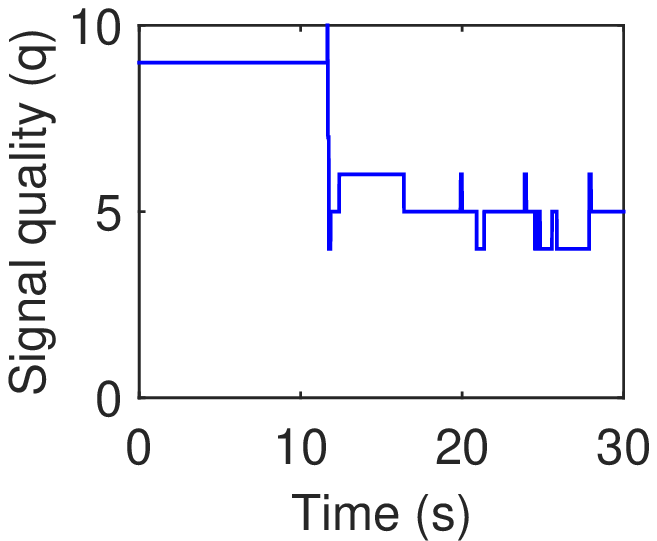}
	\label{fig3.18:scenario3_3m_a}}	
	\subfigure[Throughput, 3\,m] 
	{\includegraphics[width=0.21\textwidth]{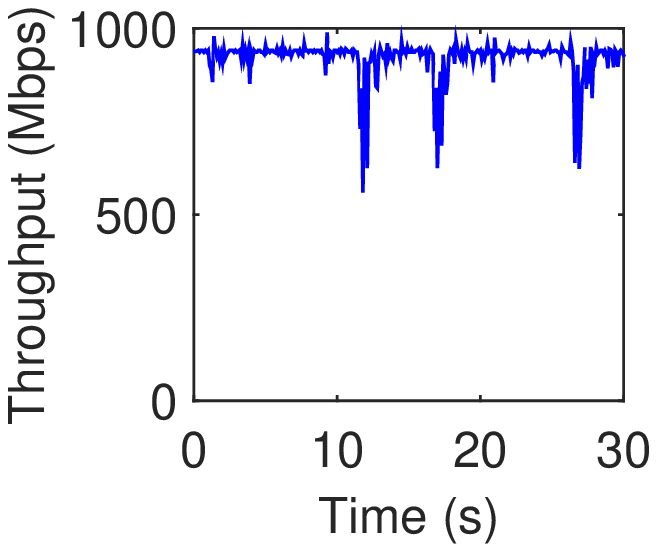}
	\label{fig3.18:scenario3_3m_b}}
	\subfigure[Signal quality, 7\,m] 
		{\includegraphics[width=0.21\textwidth]{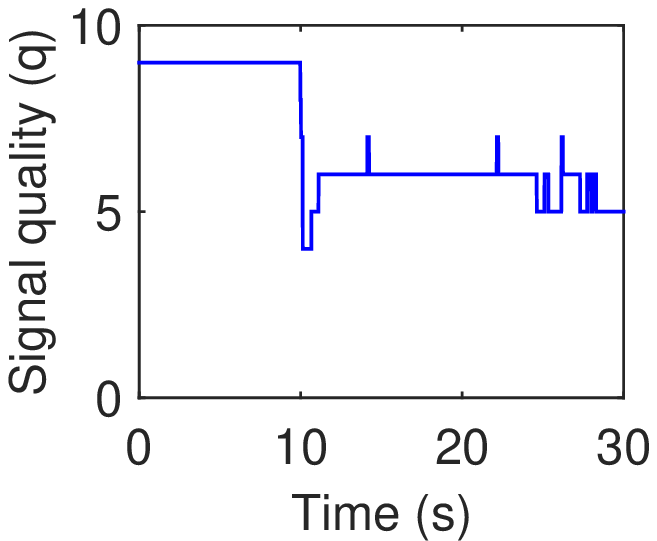}
		\label{fig3.19:scenario3_7m_a}}	
		\subfigure[Throughput, 7\,m] 
		{\includegraphics[width=0.21\textwidth]{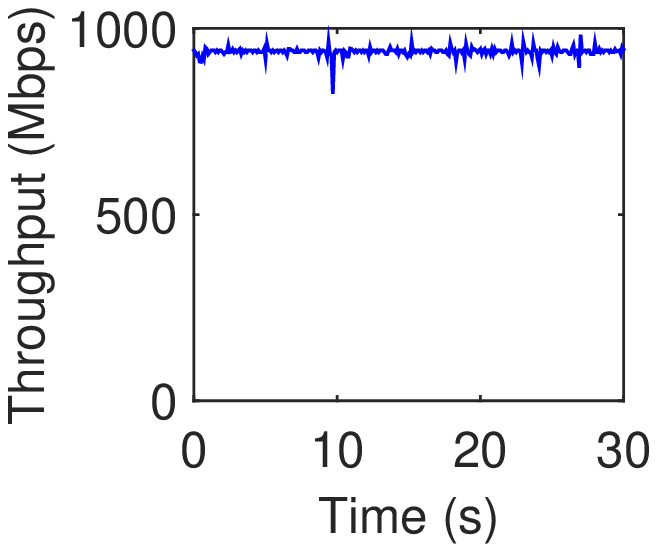}
		\label{fig3.19:scenario3_7m_b}}
		\caption{Permanent blockage with NLOS path, $d$ = 3\,m and $d$ = 7\,m.}
	\label{fig3.18:scenario3_3m}	
	\vspace{-4mm}
\end{figure}

In this scenario STA and AP were placed  1\,m away from a concrete wall.  This experiment shows how the wall acting as a reflector can be exploited to provide an alternative NLOS transmission path when the LOS path is blocked. 
%
%
For $d$ = 3\,m,  Fig.\ref{fig3.18:scenario3_3m_a} and Fig.\ref{fig3.18:scenario3_3m_b}  show the sample results for signal quality $q$ and throughput performance, respectively. When the person is in between the AP and STA, the signal quality decreases from $q=9$ to $q=4$ and fluctuates between $q=4$ and $q=6$. The throughput performance is not significantly affected by the blockage except  few small dips. The reason is that when the person obstructs the LOS transmission path, STA and AP automatically perform beam switching and take advantage of the signal reflected from the wall to establish an alternative NLOS transmission path. Since the wall is very close (1\,m) to both the devices and the wall is highly reflective, high throughput performance is maintained. 
The signal quality $q$ and throughput variations at $d$ = 7\,m are shown in Fig.~\ref{fig3.19:scenario3_7m_a} and Fig.~\ref{fig3.19:scenario3_7m_b}, respectively. We can observe that  for  $d$ = 7\,m,  fluctuations in the throughput are less as compared to that of $d$ = 3\,m. This is because, when an obstacle is close to STA/AP, human obstructing the NLOS path eclipses the whole beam transmission area.  As the distance between STA/AP and the obstacle increases, the human blocking is less likely to affect the reflected beams.   

%
Since the highest achievable data rate is limited to  1\,Gbps due to the limitations posed by the Ethernet port of the laptop hosting the iPerf server (for measurement purposes), the differences between the throughput performance in LOS and NLOS transmission path can not be observed directly. If it was possible to establish 6-7\,Gbps connection, the difference among throughput of LOS and NLOS path would be higher. Nevertheless, our experiments demonstrate that even if LOS is blocked for a long duration, a 60\,GHz link using COTS IEEE 802.11ad chipsets can resurrect itself by harnessing the reflective paths. 
\section{Handoffs  in IEEE 802.11ad WLAN}
mmWave APs/BSs will be an integral part of future ultra-dense picocellular networks and WLANs where inter-BS/AP distances could be as low as 10\,m. If the mmWave link between an STA and AP/BS is obstructed, it is highly likely that other mmWave AP/BS could be found to establish an alternate link. However, we have seen in the previous section that even if the link is blocked, the blockage can be transient in nature or an NLOS connection guaranteeing desired quality of service/experience can be established. Therefore, it is not always required to switch to an alternate AP/BS if link blockage is experienced. In the previous section, we have also seen that when $d=3\,m$, even if a permanent human blockage occurs, the link can still function. Apart from blockages, mobility is another reason for handoffs. Thus intelligent handoff techniques are highly desired to minimize the link disruption as well as to avoid the frequent association/disassociation with AP/BS. 
\subsection{Blockage induced handoff}\label{sec:Handoff}
\begin{figure}[!]
	\centering
	\includegraphics[width=0.44\textwidth]{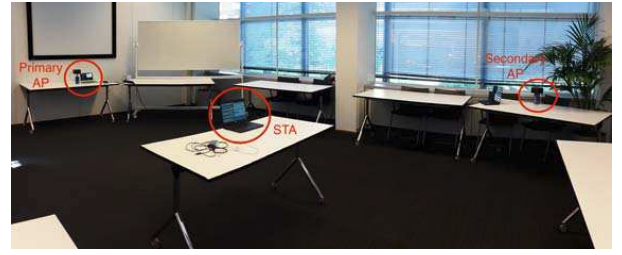}
	\caption{Blockage induced handoff measurement setup.}
	\label{fig4.5:handoff_time_evaluation_setup}
	\vspace{-4mm}
\end{figure}

In this section, we present a handoff mechanism that determines whether AP/BS switching is required or not by characterizing human blockages  into the permanent blockage and transient blockage. In the case of transient blockage, STA is not required to switch to alternate AP/BS, however, if a permanent blockage occurs, the handoff is performed by switching to other AP/BS. We use two COTS WiGig APs and one STA as shown in Fig.~\ref{fig4.5:handoff_time_evaluation_setup}.

\begin{figure}[!]
	\centering
\includegraphics[width=0.32\textwidth]{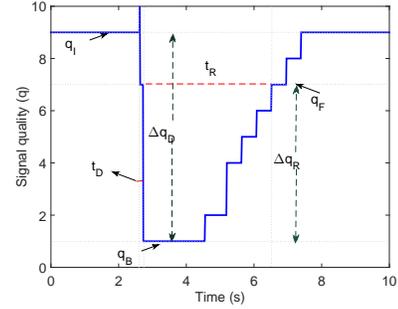}
	\caption{Signal quality drop and recovery parameters during transient blockage.}
	\label{fig4.1:sq_drop_recover_analysis}
	\vspace{-4mm}
\end{figure}

\textbf{Parameters for characterizing blockage:} Let  $q_I$ be the initial signal quality when there is no shadowing. After the link is obstructed, the lowest signal quality is denoted by $q_B$. The difference in signal quality before and during blockage is represented by  $\Delta q_{D}=q_I-q_B$ units. Let $t_D$ be the time elapsed for signal quality to drop by $\Delta q_{D}$ units.  If the blockage is transient or there is an NLOS path available,  then signal quality will recover to some value $q_f$ once the blockage disappear or an NLOS path is established. \normalcolor Lets us denote the rise in the signal quality as $\Delta q_{R}=q_F-q_B$ and the time required to reach $q_F$ as $t_R$.  Fig.~\ref{fig4.1:sq_drop_recover_analysis} shows a snapshot of the transient blockage scenario with $\Delta q_{D}$ = 8, $q_{R}$ = 6, $t_D$ = 160\,ms and $t_R$ = 3.789\,s. 

\noindent\textbf{Characterizing $t_D$ and $t_R$:} $\overline{t_D}$ is determined by calculating the difference between the time when blockage starts and the time at which the lowest value of  signal quality $q$ is observed. To determine  recovery time $\overline{t_R}$, for Scenario-1 we used $q_{R}$ = 6 as the threshold values  and for Scenario-3 we used $q_{R}$ = 2. A higher $q_{R}$ is chosen for Scenario-1 as we observed in the previous section that the drop in signal quality  $q$ is very significant in this case, and as the temporary blockage is over, the signal quality $q$ increases significantly. On the other hand, in the case of Scenario-3, drop and rise in signal quality $q$ are of small magnitudes.    $\overline{t_R}$ is not applicable to  Scenario-2 as signal recovery is negligible due to the unavailability of an alternate NLOS path. However, we noticed that there are some spikes in signal quality $q$  probably due to the ground reflections. We repeated experiments 100 times  for each of the blockage scenario to characterize $q_{D}$, $q_{R}$, $t_D$ and $t_R$.
 
Table~\ref{tab4.1:drop_recovery_time} shows the average $t_D$ and $t_R$, denoted as $\overline{t_D}$ and $\overline{t_R}$, respectively, for all the three scenarios. Ideally  $t_D$ + $t_R$ is the time duration after which blockage can be categorized. However, we see that the difference in average and maximum values of $t_D$ and $t_R$ is quite high.
\begin{table}[!]
	\centering
		\caption{ Measured $t_D$ and $t_R$ for three blockage scenarios.}
	\begin{tabular}{|l l|l|l|}
		\hline 
		&                      & $d$ = 3\,m    & $d$ = 7\,m \\
		\hline \hline
		        & $\overline{t_D}$     & 197 ms   & 140 ms \\
Transient blockage		& $t_{D_{max}}$        & 838 ms      & 513 ms \\		
		(Scenario-1)        & $\overline{t_R}$     & 3.826 s     & 1.648 s \\
		        & $t_{R_{max}}$        & 5.726 s     & 2.434 s \\
		\hline
Permanent blockage		& $\overline{t_D}$     & 232 ms    & 298 ms \\
		(Scenario-2)        & $t_{D_{max}}$        & 748 ms      & 952 ms \\ 
		        & $\overline{t_R}$     & NA     & NA \\
		        		        & $t_{R_{max}}$        & NA     & NA \\
		\hline
Permanent blockage with NLOS		& $\overline{t_D}$     & 267.65 ms   & 411 ms \\
	(Scenario-3)	        & $t_{D_{max}}$        & 707 ms      & 784 ms \\
		        & $\overline{t_R}$     & 190 ms     & 136 ms \\
		        & $t_{R_{max}}$        & 376 ms     & 313 s \\
		\hline
	\end{tabular}
	\label{tab4.1:drop_recovery_time}
	\vspace{-3mm}
\end{table}

\noindent\textbf{Characterizing $\Delta q_{D}$ and $\Delta q_{R}$:} To assess $\Delta q_{D}$ and $\Delta q_{R}$ we set threshold on $t_D$  and $t_R$ as  $t_{D_{th}}$ = 1\,s and $t_{R_{th}}$ = 3\,s based on the data from Table~\ref{tab4.1:drop_recovery_time} such that the maximum values of $t_D$ and $t_R$ are included. 

\begin{figure}[!]
	\centering
	\includegraphics[width=0.38\textwidth]{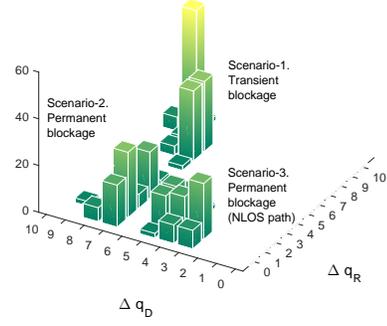}
	\caption{$\Delta q_{D}$ and $\Delta q_{R}$ for each blockage scenario.}
	\label{fig4.3:blockagecharacterization}
	\vspace{-4mm}
\end{figure}

Fig.\ref{fig4.3:blockagecharacterization} shows the histogram of $\Delta q_{D}$ and $\Delta q_{R}$ for each of the blockage scenario at $d$ = 7\,m. From Fig.~\ref{fig4.3:blockagecharacterization}, we can conclude that each blockage type has a unique $\Delta q_{D}$ and $\Delta q_{R}$ cluster. 
For example, the blockage in Scenario-1 is indicated by large $\Delta q_{D}$ and $\Delta q_{R}$, while the blockage in Scenario-2 is indicated by a large $\Delta q_{D}$ and a small $\Delta q_{R}$. On the other hand, blockage in Scenario-3, i.e., permanent blockage with the availability of NLOS path, both $\Delta q_{D}$ and $\Delta q_{R}$ have small values.

The average values of $\Delta q_{D}=x_c$ and $\Delta q_{R}=y_c$ for three blockage scenarios are shown in Table~\ref{tab4.2:detaSQ_drop_recovery}. Now with this thorough characterization of the link quality for various types of blockage, we now propose a handoff algorithm.
\begin{table}[!]
	\centering
		\caption{$\Delta q_{D}$ and $\Delta q_{R}$ measurements.}
	\begin{tabular}{|l l|l|l|}
		\hline 
		& \hspace{-3cm}Blockage scenario              & $d$ = 3\,m    & $d$ = 7\,m \\
		\hline 
		  Transient blockage	&  $\Delta q_{D}(x_c)$        & 7.5     & 7.72\\		
		(Scenario-1)        & $\Delta q_{R}(y_c)$     & 7.32      &7.64 \\
		     
		\hline
Permanent blockage		&  $\Delta q_{D}(x_c)$     & 7.20    & 7.30 \\
		(Scenario-2)        & $\Delta q_{R}(y_c)$        & 1.41      & 1.56 \\ 
		       		\hline
Permanent blockage with NLOS		&  $\Delta q_{D}(x_c)$     & 3.40   & 4.06 \\
	(Scenario-3)	        & $\Delta q_{R}(y_c)$        & 1.13      & 1.20 \\
		       		\hline
	\end{tabular}
	\label{tab4.2:detaSQ_drop_recovery}
	\vspace{-3mm}
\end{table}

\noindent\textbf{Handoff algorithm:}  The Euclidean Distance ($ED$) between the measured $\Delta q_{D}$,  $\Delta q_{R}$  defined as $x$ and $y$ and the average $x_c$ and $y_c$ is defined as, 
\begin{equation} \label{eq4.1:euclididan_distance}
ED_i = \sqrt{(x - x_{c,i})^2 + (y - y_{c,i})^2}.
\end{equation}
Where $i\in\{1,2,3\}$ represent the three blockage scenarios, respectively.

The smallest $ED_i$ implies that the measured changes  in the signal quality $q$ (i.e., $\Delta q_{D}$ and $\Delta q_{R}$) indicate that the $i^{th}$ blockage scenario has occurred. The blockage characterization algorithm is described in Algorithm-\ref{pseudocode} and explained in detail as follows.
\begin{algorithm}
\caption{Handoff mechanism based on the human blockage characterization.}\label{pseudocode}
\begin{algorithmic}[1]

\WHILE {($t \leq t_{D_{th}}$)}
	\STATE monitor $\Delta q_{D}$ \\
	\hspace{1.3em} $x \gets \max(\Delta q_{D}$) 
\ENDWHILE 

\IF {($x > 2$)}
	\STATE Blockage indication
	\WHILE {($t \leq t_{R_{th}}$)}		
		\STATE monitor $\Delta q_{R}$ \\
		\hspace{2.8em} $y \gets \max(\Delta q_{R}$) 
	\ENDWHILE	
	\STATE calculate $ED_1, ED_2, ED_3$	\\		
	\hspace{1.3em} $b \gets \min(ED_1, ED_2, ED_3)$ 
		
	\IF {($b = ED_1 \hspace{0.5em}$ or $\hspace{0.5em} b = ED_3)$}
	\STATE Short-term blockage 
	\ELSE
	\STATE Long-term blockage 	
	\ENDIF	
\ELSE
	\STATE No blockage
\ENDIF
\end{algorithmic}
\end{algorithm}
During the pre-defined $t_{D_{th}}$, a STA monitors  $\Delta q_{D}$ at every $t$ = 1\,ms interval. If the maximum $\Delta q_{D}$ exceeds 2, then there is a blockage indication. Otherwise, there is no blockage indication. We set the $\Delta q_{D}$ threshold to 2 but not 1 because there are always some small variations in signal quality that may result in a difference of 1 unit.
 
After detecting the blockage indication, STA monitors the $\Delta q_{R}$ during the predefined $t_{R_{th}}$. The maximum $\Delta q_{D}$ and $\Delta q_{R}$ is then defined as $x$ and $y$, respectively.  After getting the $x$ and $y$ values, $ED_1$, $ED_2$ and $ED_3$ are calculated. The minimum distance among $ED_1$, $ED_2$ and $ED_3$ is indicated as the corresponding blockage type, referred to as $b$. If $b$ equals to either $ED_1$ or $ED_3$, then STA infers that a short-term blockage has happened. Otherwise, a long-term blockage is indicated which is followed by handoff initiation.  
 
\begin{figure}[!]
 	\centering 
 	\subfigure[Signal quality.] 
{\includegraphics[width=0.22\textwidth]{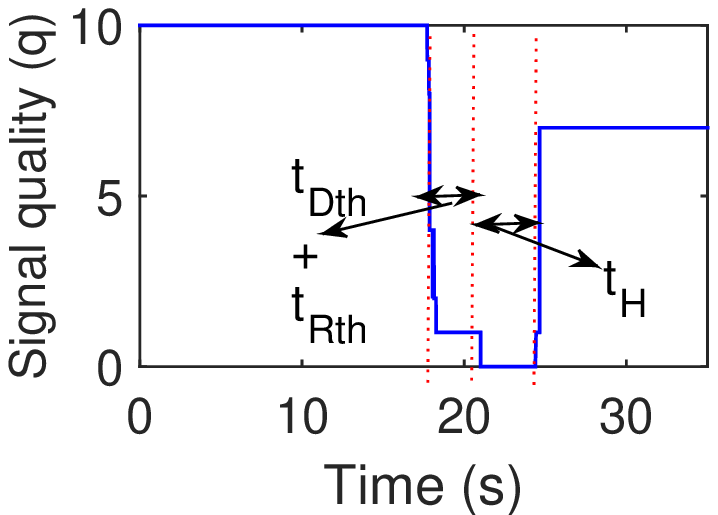}
 	\label{fig4.6:handover_performance_a}	}	
 	\subfigure[Throughput.] 
 		{\includegraphics[width=0.22\textwidth]{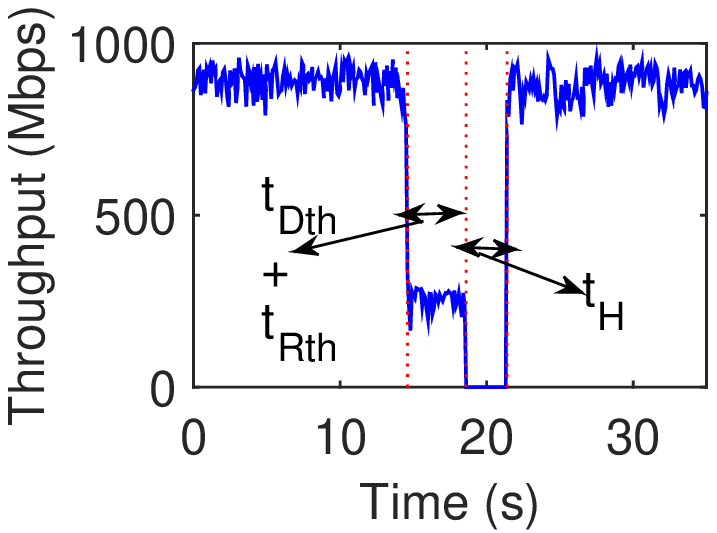}
 	\label{fig4.6:handover_performance_b}	}
 	
 	\caption{Signal quality and throughput during handoff procedure.}
 	\label{fig4.6:handover_performance}	
 	\vspace{-3mm}
 \end{figure}

Fig.~\ref{fig4.6:handover_performance_a}  and Fig.~\ref{fig4.6:handover_performance_b} demonstrate  the signal quality and throughput variation during the handoff triggered due to a permanent blockage. The total time taken in a blockage induced handoff procedure consist of $t_{D_{th}}$ + $t_{R_{th}}$ +$t_H$.  From our handoff experiments  we found that average $t_{H}$ is \textbf{2.749\,s}. Since we used $t_{D_{th}}$=1\,s and  $t_{R_{th}}$=2\,s, the average  time to switch the AP is 6.749\,s.

\noindent\textbf{Accuracy of blockage detection:} We observe that the proposed blockage characterization relying on the fluctuations in signal quality parameters ($\Delta q_{D}, \Delta q_{R}$) mainly depends on the threshold observation durations $t_{D_{th}}$ and $t_{R_{th}}$. Hence, its important to determine the accuracy of characterization. We define $A_{T}$ and $A_{P}$ as the accuracy of detecting transient blockage and permanent blockage, respectively.  To assess the impact of $t_{D_{th}}$ and $t_{R_{th}}$ on $A_{T}$ and $A_{P}$, performed experiments considering multiple $t_{D_{th}}$ and $t_{R_{th}}$.\normalcolor We performed 50 measurements for each combination of $t_{D_{th}}$ and $t_{R_{th}}$ for each of the blockage type.
\begin{table}[!]
	\centering
	\caption{Detection accuracy for various $t_{D_{th}}$.}
	\begin{tabular}{|c|c|c|c|}
		\hline 
		$t_{D_{th}}$ & $t_{R_{th}}$ & $A_{P}$ & $A_{T}$ \\
		\hline \hline
		250\,ms &		&  72\,\% &  56\,\%  \\	
		500\,ms & 1\,s	&  76\,\% &  66\,\%  \\ 
		750\,ms & 	    &  88\,\% &  74\,\%  \\ 
		\hline		
		250\,ms &		&  96\,\% &  73\,\%  \\
		\textbf{500\,ms} & \textbf{3\,s}	& \textbf{97\,\%} & \textbf{96\,\%}  \\
		750\,ms & 	    &  99\,\% & 97\,\%  \\
		\hline
		250\,ms &		& 100\,\% &  72\,\%  \\
		500\,ms & 6\,s	&  96\,\% &  92\,\%  \\
		750\,ms & 	    & 100\,\% &  96\,\%  \\
		\hline				
	\end{tabular}
		\label{tab5.2:accuracy_tdrop}
\end{table}
Table~\ref{tab5.2:accuracy_tdrop} shows the accuracy of blockage type detection for different values of $t_{D_{th}}$ and $t_{R_{th}}$. We can observe that when  $t_{Rth}$=1\,s  false detection is particularly high for the transient blockage scenario. This is because the  transient blockage is characterized by  a high $\Delta q_{D}$ as well as a high $\Delta q_{R}$. However, a smaller recovery time threshold may not allow enough rise in signal quality $q$. This results in interpreting transient blockage as the permanent one. The accuracy in detection of permanent blockage is generally high. As $t_{R_{th}}$ and $t_{D_{th}}$ increases accuracy of  both blockage types detection increase. A large $t_{R_{th}}$ + $t_{D_{th}}$  means delay in handoff decision which can result in a low QoS/QoE if link is obstructed by the permanent blockage. It appears that $t_{R_{th}}$=500\,ms and $t_{D_{th}}$=3\,s  provide reasonably high accuracy.
\subsection{Mobility induced handoff in indoors} One of the important characteristics of 60\,GHz based picocells in an indoor environment is the limited overlap area between neighboring APs which is also called corner effect. This is one of the main hurdle in providing a WiFi-like seamless indoor coverage at 60\,GHz frequency band.
\begin{figure}[!]
	\centering 
	\includegraphics[width = 0.28\textwidth]{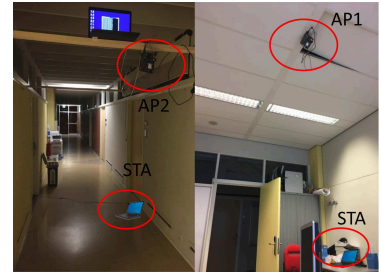}
	\caption{Experimental scenario for handoff triggered by corner effects.}
	\label{FigAPceilingwithlaptop}	
	\vspace{-4mm}
\end{figure}

\begin{figure}[!]
 	\centering 
 	\subfigure[Signal quality.] 
 	{\includegraphics[width=0.20\textwidth]{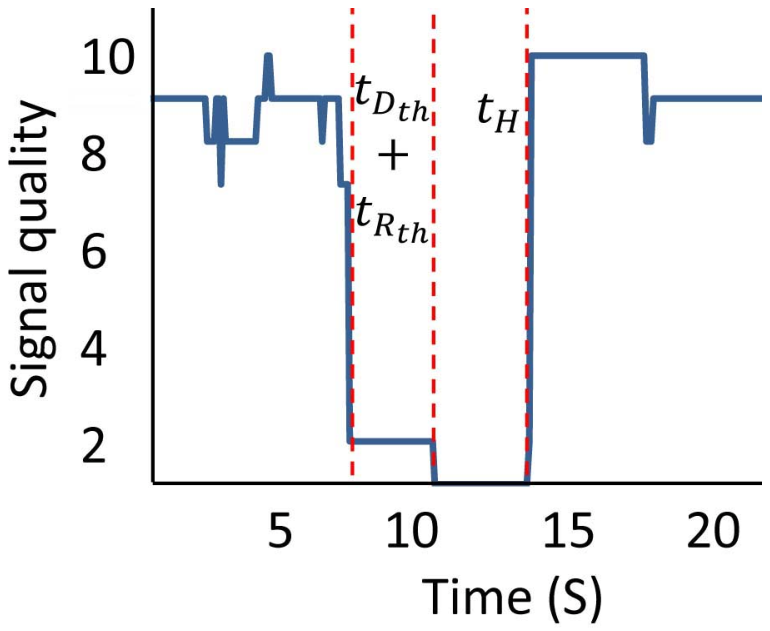}
 	\label{corner1}	}	
 	\subfigure[Throughput] 
 	{\includegraphics[width=0.20\textwidth]{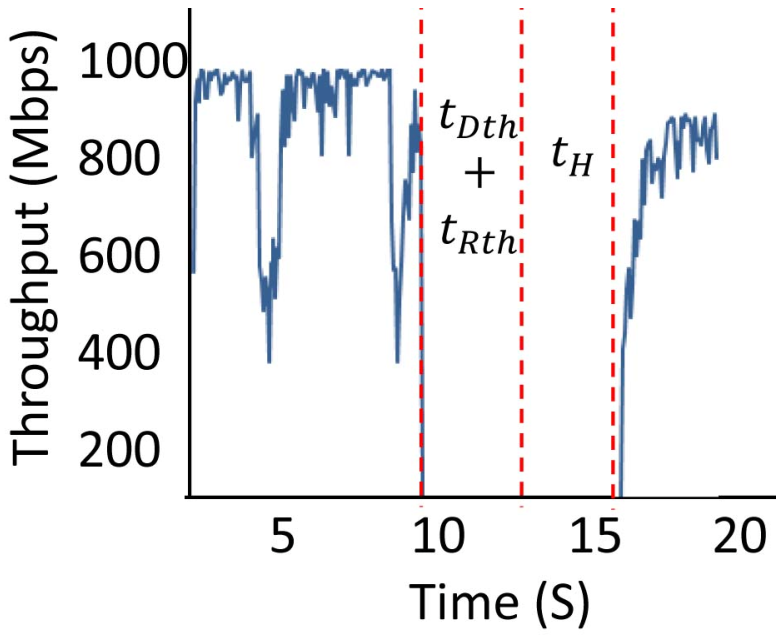}
 	\label{corner2}	}
 	
 	\caption{Handoff between AP located in the room and in the corridor.}
 	\label{corner_performance}	
 	\vspace{-4mm}
 \end{figure}

 Fig.~\ref{FigAPceilingwithlaptop} shows such a scenario we used for our experiments. AP1 is placed at the ceiling of office room while AP2 is in the ceiling of corridor. Inititailly STA is in the room and connected with AP1. Fig.~\ref{corner_performance} shows the signal quality $q$ and corresponding iPerf throughput variation when the STA connected with  AP1  walks out of the room. As soon as STA crosses the door and reaches behind the wall, a sudden dip in the signal quality $q$ and throughput is observed, although the traveled distance is only 2-2.5\,m. This phenomenon is not observed in 2.4/5\,GHz as signals can easily penetrate walls. For our handover mechanism, as soon the signal quality falls below $q=2$, STA disassociates itself (assuming a permanent blockage) with  AP1 and is able to connect with the AP2 located in the corridor in 3\,s. 
\subsection{Comments on the measured handoff delay}
Generally, when an STA moves away from the connected  AP, the signal quality decreases. In traditional IEEE 802.11 WLANs, if signal quality decreases below a certain threshold, the STA sends the probe packets to confirm the presence of other APs. The exchange of probe request and response messages constitutes the discovery phase. Once the STA has discovered an alternate AP, authentication and association procedures are followed. Let the delays incurred in discovery, authentication and association be represented by $t_{dis}$, $t_{auth}$ and $t_{as}$, respectively. In the proposed  IEEE 802.11ad handoff mechanism, the additional delay in making handoff decision (due to the characterization of blockage type), i.e., $t_{R_{th}}$+$t_{D_{th}}$ is also involved.\normalcolor Thus the total delay  $t_s$ incurred in switching from one AP to another AP can be defined as,
\begin{equation}
t_s=t_{R_{th}}+t_{D_{th}} +t_{dis}+t_{auth}+t_{as}.
\end{equation}
Here, $t_H=t_{dis}+t_{auth}+t_{as}$ as shown in Fig.~\ref{corner_performance}.
The measured average switching time in our experiments is  6.749\,s which is quite high  compared to the legacy IEEE 802.11 WLANs where it generally takes less than a second in handoffs. The main reasons behind fast handoffs in WiFi are: (i)~WiFi STAs always keep a list of alternate  APs which reduces the discovery delay; (ii)~inter-AP connectivity in WiFi networks greatly reduces the association and authentication delay; and (iii)~blockage characterization and beam-searching are not needed. An important reason behind the high handoff delay could be that the docking stations we used in our experiments are not optimized for WLANs operations.\normalcolor To provide seamless multi-Gbps wireless connectivity, smooth handoffs are a must for next generation of 60\,GHz systems. This requires better network design approaches particularly considering the above listed factors and a possibly close-integration with the sub-6\,GHz air-interfaces.
 \section{Conclusion}\label{sec:conclusion}
 Among millimeter wave (mmWave) frequencies, 60\,GHz band is leading in terms of research efforts and the availability of commercial devices. The inevitability of mmWave communications in future networks warrants careful performance evaluation of commercial off-the-shelf (COTS) 60\,GHz devices. This paper presents an experimentation driven extensive measurements based work. We carefully designed scenarios and through thorough practical evaluation, we provided important insights about the directional antenna patterns, alignment and association overheads and interference and blockage characteristics of COTS IEEE 802.11ad devices from a WLAN/picocell perspective where frequent handoffs can be triggered. Through multiple measurements, we have provided a highly accurate (96-97\%) blockage characterization mechanism to assist link blockage induced handoffs.  Our in-depth analysis of human blockage shows that it is important to differentiate between the blockage types to avoid unnecessary handoffs. Our insights, backed by accurate quantitative measurement data, is highly beneficial in designing and improving the efficiency of next generation mmWave networks. The detailed experimentation results provided in this work can act as a benchmark for future design of indoor 60\,GHz picocells.
\bibliographystyle{IEEEtran}
\bibliography{ReferencesmmNets}
\end{document}